\documentclass[conference]{article}

\usepackage{graphicx}
\usepackage[table,xcdraw]{xcolor}
\usepackage[hyphens]{url}
\usepackage{breakurl}
\usepackage[hidelinks]{hyperref}
\hypersetup{breaklinks=true}

\usepackage[color=yellow!60,textsize=footnotesize,obeyDraft,draft]{todonotes}
\usepackage{enumitem}
\usepackage{subcaption}
\usepackage{caption}
\usepackage{amssymb}
\usepackage{amsmath}
\usepackage{amsthm}
\usepackage{cleveref}

\usepackage{placeins}

\usepackage{cite}

\usepackage{graphicx}

\begin{document}
\captionsetup[figure]{font={scriptsize,sf},labelformat={default},name={Fig.},labelsep=period}

\title{Accurate Energy Modelling on the Cortex-M0 Processor for Profiling and Static Analysis}

\author{Kris Nikov, Kyriakos Georgiou, Zbigniew Chamski,\\ Kerstin Eder\thanks{K. Nikov, K. Georgiou, Z. Chamski and K. Eder are with the University of Bristol, UK.
(e-mail: firstname.lastname@bristol.ac.uk)} 
  { }and  Jose Nunez-Yanez%
\thanks{J. Nunez-Yanez is with Linköping University, Sweden.
(e-mail: jose.nunez-yanez@liu.se)}
}%

% make the title area
\maketitle

\begin{abstract}

%Edge computing enables more processing on the spot and thus, reduces the interaction with the cloud. While this results in reduced latency and bandwidth utilization and increased quality of IoT services, it adds more pressure to the software developer as edge computing devices have limited resources, such as compute power and energy budgets. 
%
Energy modelling can enable energy-aware software development and assist the developer in meeting an application's energy budget. Although many energy models for embedded processors exist, most do not account for processor-specific configurations, neither are they suitable for static energy consumption estimation. 
This paper introduces a set of comprehensive energy models for Arm's Cortex-M0 processor, ready to support energy-aware development of edge computing applications using either profiling- or static-analysis-based energy consumption estimation. We use a commercially representative physical platform together with a custom modified Instruction Set Simulator to obtain the physical data and system state markers used to generate the models. The models account for different processor configurations which all have a significant impact on the execution time and energy consumption of edge computing applications. Unlike existing works, which target a very limited set of applications, all developed models are generated and validated using a very wide range of benchmarks from a variety of emerging IoT application areas, including machine learning and have a prediction error of less than 5\%.

\end{abstract}

\section{Introduction}
\label{sec:intro}

One trillion new Internet of Things (IoT) devices are predicted to reach the market by 2035~\cite{ARM_trillion_iot_devices} ushered by the increasingly expanding edge computing market.
% These devices would be generating an unprecedented amount of data that would need to be pushed to the cloud for storing and processing~\cite{IDC_iot_data_2025}.
% The emergence of 
%Edge computing, however, has enabled a degree of processing at the data-source that avoids the need for transmitting all collected data to the cloud. Although this can significantly reduce response time and bandwidth requirements, it results in increased resource requirements from edge devices, such as processing power and energy. 
Typically, IoT devices are not part of a power grid but rather are scattered in the environment and powered by limited energy sources, such as batteries or energy harvesting. Thus, they are mostly based on small embedded processors with a tiny energy footprint, such as the Arm Cortex-M0. This kind of processor is inherently limited in processing power, making edge computing challenging. Developers must apply extreme optimisations to trim down the processing time, memory, and energy consumption of algorithms to enable their execution on such small embedded devices. A trending example is the streaming down of traditional machine learning algorithms to enable their execution on tiny IoT devices~\cite{ARM_Cortex_M_machine_learning_white_paper}. 

The burden now lies with the software engineers to develop edge computing applications that can fit on the limited memory of the IoT embedded devices, execute within reasonable timeframes,
% depending on the type of the application,
and run within the available energy budget. %Balancing the resource usage of an IoT application is a challenging act and typically a manual trial-and-error process that costs significant development time. Thus, practical methodologies that enable resource-aware software development will significantly help to tackle the challenges associated with the development of edge computing applications. 
Execution time and code size are easy to measure and well understood by the typical software developer. On the contrary, energy consumption information is not readily accessible, and something most software developers never had to account for. For edge computing, however, energy consumption feedback during the applications' development cycle is at least equally important as execution time and code size%. 
%
% {\color{red}Many will argue that execution time alone is sufficient, considering energy consumption as a proxy of time, but this is not always the case.}
%
%Even when the energy consumption is directly proportional to time, energy consumption figures are still needed to ensure that the application's available energy budget is met, therefore development tools need to enable such feedback
~\cite{GeorgiouK:2020,Georgiou:2017}. 
%or to counterbalance between the energy consumption and execution time whenever there is a choice 

Hardware measurements are the most accurate way of acquiring a program's energy consumption information, but they are not broadly supported by the hardware vendors and not within the know-how of typical software developers. Energy modelling and the integration of energy models into the development toolchains can solve both of these issues~\cite{Georgiou:2017}. Once an accurate energy model has been developed for a particular platform, it can be integrated into a toolchain to allow for energy estimations with each compilation. 

The literature offers a plethora of energy consumption models for embedded processors~\cite{Penolazzi:2009,Brandolese2011,nikov2020intra,nikov2022robust,Yassin2020}. For an energy model to be useful to the software developer, it must be able to convey energy consumption information at the source-code level. Thus, Instruction-Set-Architecture-based (ISA) energy models~\cite{Tiwari1996} became the most popular, because modelling at the ISA level allows for attributing energy costs to software components, such as ISA Control Flow Graph (CFG) basic blocks. %Therefore, ISA-based energy models can be utilized by compiler autotuning techniques to discover energy targeted compiler optimizations~\cite{GeorgiouK:2020}, and by static analysis tools~\cite{Georgiou:2017} that estimate the energy consumption of programs. 
Although ISA-based energy modelling approaches have benefits, extracting such models is time-consuming and challenging. It requires devising often complex energy measuring procedures to capture the energy consumption of each instruction in the ISA. %Typically, an instruction is executed in a tight loop while measuring the power dissipated together with the execution time. For instructions that can not be measured within such a loop, for example branch instructions, regression analysis is needed to capture their energy consumption. 
On the other hand, energy modelling using Performance Monitoring Counters (PMCs), also named hardware event counters, is a more accessible approach compared to ISA-level modelling. It requires measuring the energy consumption of representative programs, collecting execution statistics from PMCs and then deducting energy consumption coefficients via mathematical analysis and machine learning techniques~\cite{nikov2020intra,nunez2020run,nikov2022robust}.

This paper demonstrates how to build PMC-based models for multiple embedded-processor configurations. The models can be used to attribute energy costs to software components and facilitate both profiling-based and static-analysis-based energy consumption estimation, similar to ISA-based models. Our main contributions are:

\begin{enumerate}
\item Due to limited support for PMCs on most IoT platforms, we customised an open-source Instruction Set Simulator (ISS) of the Arm \texttt{thumb} ISA, namely the \texttt{Thumbulator}~\cite{Thumbulator} to produce accurate execution statistics useful for developing energy models.
\item We identified a set of PMCs that are both statically predictable at ISA basic block level and offer an energy consumption estimation error (a Mean Absolute Percentage Error (MAPE) of less than 5\%)~\cite{discovery-data}.
%
%\end{enumerate}

%Only a few of the existing ISA models for deeply embedded processors are parametrized by hardware configurations, but only cover the processor's frequency and voltage~\cite{Yassin2020}. Other processor configurations, such as instruction buffer configurations, are equally important as they have a major impact on both the execution time and the energy consumption of an embedded processor. 

%\begin{enumerate}\setcounter{enumi}{2}
\item We enhanced \texttt{Thumbulator} to include advanced configurations for the STM32F0xx family of processors~\cite{STMFO_technical}. We tracked the use of the instruction PreFetch buffer (ON/OFF) which increases the efficiency of instruction fetching and the number of CPU WaitStates (0/1) %\todo[inline]{Terms like Waitstate are not standard terms in research and could be defined in short.} 
 required to correctly perform read operations from Flash memory (mandatory at higher CPU frequencies since flash memory latency is higher than the CPU clock speed).% so the CPU requires additional time to wait for the data to arrive from memory.
%
%Our energy models include all the permitted combinations of the selected configurations for a set of commonly used processor frequencies: 20, 24, and 48 MHz. These energy models allow software developers to select the configuration that will meet their application's energy consumption and execution time goals. %They can also potentially be used to assess the application's risk of exposure to side-channel attacks (see~\Cref{subsec:security}).
\end{enumerate}

\section{Energy Modelling Methodology}
\label{subs:OptRealApps}

\subsection{Measurement Setup}

\begin{figure}[!htp]
    \centering  
    \includegraphics[width=\linewidth]{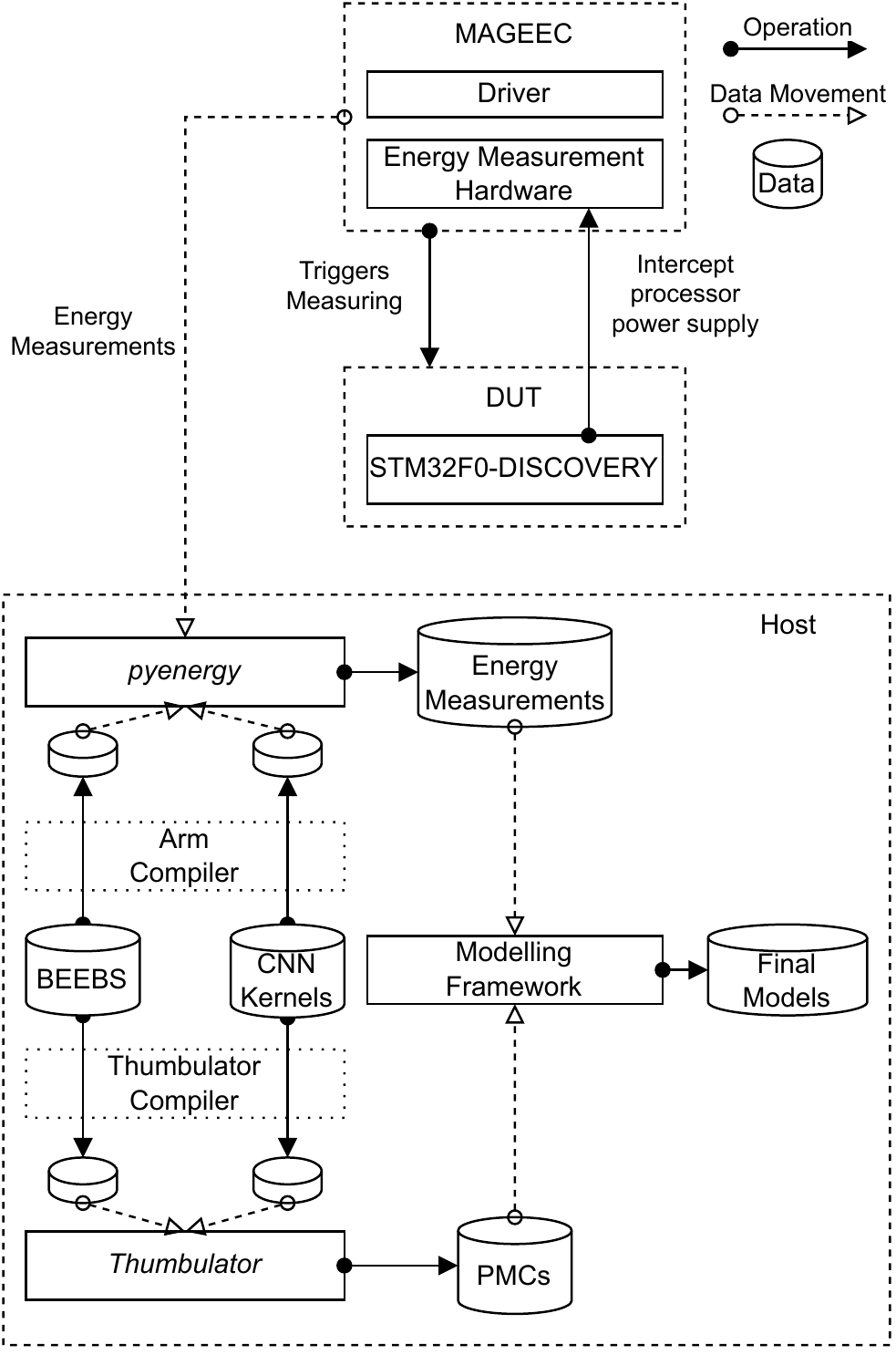}
    \caption{Hardware and software harness for energy modelling of the Cortex-M0.}
    \label{fig:cortexM0_energy_modelling_setup}
\end{figure}

Our proposed methodology involves a custom measurement set-up to extract energy consumption information from our target platform - the STM32F0-Discovery board a.k.a. the device under test (DUT) - and collating the data with PMC information from \texttt{Thumbulator} to obtain the full data used to generate the models. A diagram of the full set-up including the host PC and different components is presented in Figure~\ref{fig:cortexM0_energy_modelling_setup}.

%\begin{figure}[!htp]
%\centering
%\resizebox{0.5\textwidth}{!}{%
%  \includegraphics[width=0.5\linewidth]{Mageec_energy_measurement_setup}
%}
%%\caption{Picture of the platform physical measurements setup with the STM32F0-DICOVERY on the right and the MAGEEC board setup on the left (connected to its host board the STM32F4-DISCOVERY).}
%\caption{The MAGEEC energy measurement setup connected with the DUT.}
%\label{fig:platform_setup}
%\end{figure}

%Figure~\ref{fig:platform_setup} showcases how the measurement setup with the MAGEEC shield connects to the STM32F0-Discovery platform under test. 
We have used a custom measurement board, called MAGEEC\cite{MageecMeasuring}, to intercept and sample the CPU power supply rails of the DUT. The samples are collected at a frequency of 10kHz and then converted to digital values. % using the 12bit on-board ADCs. 
 All of this is controlled via a python module called \textit{pyenergy}, which is also used to flash and run the pre-compiled workloads on the target device. The workloads are compiled for bare-metal execution using \textit{GNU-GCC}. The \textit{pyenergy} control program runs on a host platform, connected to the measurement set-up via USB. All the physical DUT measurements are saved back on the host device as a series of .csv files. 
 
 The PMCs used for platform state characterisation and model generation are obtained using \textit{Thumbulator}. The simulator has been modified to closely match the execution profile and memory set-up of the DUT. Further details about the modifications and the resulting accuracy are presented in Subsection~\ref{subsec:cm0_thumbulator_pmc}. The two sets of binaries are required because the simulator does not fully handle access to 
%the peripheral devices of the STM32F0-DISCOVERY evaluation board
off-core peripherals, e.g., PLL clock generators; these should be skipped in \texttt{Thumbulator} binaries. However, the same location and alignment of benchmark code for both types of binaries was maintained.

%Both the BEEBS and CNN-based benchmarks have been compiled into two kinds of binaries. First, the benchmarks have been compiled for the STM32F0-DISCOVERY board in order to conduct energy consumption measurements. The hardware measured energy consumption of the programs provide the data for the dependent variables of our regression analysis. Second, the benchmarks have been compiled for the modified \texttt{Thumbulator} ISS in order to derive events counter values. The counter values provide the data for the independent variables of our regression analysis.  

The aim of this work is to develop an accurate CPU model using ISS information, therefore the DUT peripherals and their interaction with the CPU are not included in the energy measurement collection and simulation. Whole-system modelling is a very important topic, especially for embedded devices and IoT and remains an area for future research.

\subsection{Benchmark Selection}
Two sets of benchmarks were used for model characterisation and validation. First, the BEEBS benchmark suite~\cite{pallister2013beebs}; an open-source embedded-system benchmark suite designed for exploring the performance and energy consumption characteristics of embedded architectures. It features several categories of benchmarks, selected to represent real-world application areas such as Automotive, Consumer and Security. %BEEBS supports the MAGEEC open-source energy measurement framework \cite{MageecMeasuring}. 
%The framework provides triggers to start and end the measurements and a calibration factor for each benchmark. This ensures that the benchmark is repeatedly executed in a loop until an adequate sampling number is achieved, and thus, the measurements can be trusted.
76 out of the 88 BEEBS benchmarks have been used. The remaining twelve do not fit in the available memory of our STM32F051 target chip on the DUT. The selected benchmarks have a measured energy Coefficient of Variability (CoV) of 2.61, which shows very high heterogeneity. 
%
%\todo[inline]{On CNN based frameworks, are coefficients directly read from ROM memory, or loaded into RAM memory? How does this affect the energy (and would that affect the model estimation? There is a great difference between reads from ROM (novel NVM, fused ROM or Flash and between RAM) and RAM. The different CNN benchmarks will greatly depend on it -> I don't know how the binaries are flashed onto the device and can't asnwer this :(.}
The second set of benchmarks is based on an industrial edge computing application, developed by Irida Labs~\cite{IRIDA_LABS}. The application uses a Convolutional Neural Network (CNN) and implements a smart monitoring system that can monitor, in real-time, a car parking lot with multiple parking slots to determine whether a slot is occupied or not. The different layers of the CNN, namely Convolutional, MaxPool, and Full-Connected, were isolated and configured with different hyper-parameters and optimisations, resulting in 154 distinct benchmarks, with a measured energy CoV of 1.31 indicating the diverse nature of the different CNN layers. %The MAGEEC energy measurement framework was adapted to support the CNN benchmarks. 
Overall, a total of 230 benchmarks were used for the training and validation of our energy model with a measured energy CoV of 3.41, further highlighting the diverse profile of the workload set. This number goes significantly beyond the average number of used benchmarks reported for existing energy models of embedded processors~\cite{bazzaz2013accurate,ruberg2015microcontroller,konstantakos2008energy}. %(see~\cite{ke_2015}, pages 22-23, Table 5.1). 
In order to avoid over-fitting the model, we use 10-fold cross-validation to evaluate the model performance across a variety of workload configurations. Further details on model training are available in Subsection~\ref{subsec:Cortex_m0_model_training}.

\subsection{PMC-based Code-level Energy Modelling}

PMC-based energy consumption estimation models are typically obtained via multi-linear regression analysis, where coefficients, $\beta_x$, are determined for each counter, $C_x$, to predict the overall energy cost, i.e., $E = \sum_x (\beta_x \times C_x) + \alpha$, with $\alpha$ being the residual error term. The coefficients $\beta_x$ are the constants in the energy model that are program independent while the counters $C_x$ are the variables that depend on the program and its input. For a specific program with known counters, the energy model can be used to estimate the energy consumed during the program's execution.

For static-analysis-based energy consumption estimation, the overall energy consumption estimate of a piece of code is typically constructed from the estimates of the ISA basic blocks of the program~\cite{Georgiou:2017}. Thus, %appropriate energy consumption models are needed that are capable of providing the basic blocks energy costs. 
a PMC-based energy model can enable energy consumption estimation via static analysis only if the counters used for the modelling and prediction can be statically predicted at the ISA basic block level. %Thus, a big part of our modeling efforts was on finding event-counters that are both suitable for static analysis and yet offer a low estimation error.

%Note that the static analysis cannot derive upper bounds of the counters independently for the whole program, as then the path analysis may select multiple contradicting paths for the different counters, rendering the overall result useless.

In order to make the model scalable for block-level static analysis we have trained without using an intercept, so the residual is absorbed into the other event weights. This means that at time 0 the energy predicted is zero. We have also used a Non-Negative Least Squares (NNLS) solver to guarantee positive weights for all the events in the final model, thus always guaranteeing predictable energy consumption values from the model at discreet time slices.

\subsection{Collection of Cortex-M0 Event Counters}
\label{subsec:cm0_thumbulator_pmc}

\begin{table}[!htb]
\centering
%\resizebox{\textwidth}{!}{%
\footnotesize
\begin{tabular}{|c|l|}
\hline
\multicolumn{1}{|c|}{\textbf{Counter}} & \multicolumn{1}{c|}{\textbf{Description}} \\ \hline
$C_{\text{1}}$ & Executed instructions (no Muls) \\ \hline
$C_{\text{2}}$ & Multiplication instructions - Muls\\ \hline
$C_{\text{3}}$ & Taken branches \\ \hline
$C_{\text{4}}$ & RAM data reads \\ \hline
$C_{\text{5}}$ & RAM writes \\ \hline
$C_{\text{6}}$ & Flash data reads \\ \hline
%Unused & RAM insn reads \\ \hline    % Zbigniew: Removed RAM instruction reads since no code was mapped to RAM
%Unused & Flash insn reads \\ \hline
%Unused & Flash writes \\ \hline % not sensible 
%Unused & BL insns \\ \hline
%Unused & BL word-aligned \\ \hline
%Unused & PUSH/POP PC/LR \\ \hline
\end{tabular}%
%}
\caption{Statically predictable PMCs for energy-modelling.}
\label{tab:Cortex_M0_counters}
\end{table}

%Traditionally, there are two ways to collect execution statistics for an architecture. One is to collect them directly via PMCs while executing a program on the actual architecture, provided it offers PMCs. For architectures without PMCs, the second way is via an ISS, preferably cycle-accurate. The ISS simulates the execution of a program for a specific architecture, and thus, it can collect PMCs. Since t
The Cortex-M0 is a deeply embedded architecture with minimal resources available on-chip and it does not expose any PMCs. Thus, we modified an open-source ISS, namely \texttt{Thumbulator}~\cite{Thumbulator}, to extract the necessary event counters for our energy consumption modelling. The modifications wrt.\ the reference \texttt{Thumbulator} implementation~\cite{Thumbulator} included four key aspects:

\begin{itemize}
\item Adaptation to reflect the memory organisation as well as the instruction fetch mechanism used in the STM32F0xx processor family.
\item Implementation of a range of event counters and the associated reporting mechanism.
\item Calibration and improvement of the timing behaviour of the simulation to match the hardware's behaviour.
\end{itemize}

The modified simulator can be used to simulate any of the processors in the STM32F0xx family~\cite{STMFO_technical} and can collect a large number of event counters that represent various aspects of the architecture's runtime behaviour such as the effective RAM and Flash memory accesses, taken branches, per-opcode instruction execution statistics, and interactions between instruction- and data-related memory accesses. %The timing behaviour validation of the modified \texttt{Thumbulator} against the actual hardware, using all the benchmarks, exposed a correctness bug in the implementation of the \texttt{ASR} instruction in the original \texttt{Thumbulator} code and identified a case of incorrect memory access counting. Both problems have been fixed in the version used to build the final energy model. 
%\todo[inline]{WaitStates of 0/1 is specific to STM. Other vendors might offer higher wait-states too. Similarly, three frequency options is specific and different vendors might offer more choices. This paper assumes Flash memory to be read-only, but some microcontroller vendors like TI supports Flash writes which consume a very large amount of power and might become an important PMC. Can the proposed approach be generalized/extended to such variety of options supported by different vendors? It would be good to have a short discussion/evaluation on applicability of the proposed approach for such variations.}
The execution time model derived from event counts reported by \texttt{Thumbulator} is fully cycle-accurate wrt.\ hardware execution when the instruction PreFetch buffer is disabled or the WaitState count is 0. When the PreFetch buffer is enabled and the WaitState count is 1, the MAPE of the \texttt{Thumbulator}-based timing prediction is 1.55\%. Theoretically this approach of using an ISS can be applied to other vendors or microprocessors, particularly where there is more available documentation about the micro-architectural implementation. This would allow even finer and quicker tuning of the ISS to match the DUT hardware.

%\todo[inline]{A major part of the second listed contribution is identifying the PMCs to use in the model. However no description of how this was done or method is presented, other than "Using the available architecture documentation and a series of modeling cycles". For example, how were the events with the most significant energy impact identified? Why are six events used?}
%\todo[inline]{The authors mention identification of PMCs as one of the contributions of the paper, but do not reveal the details. It is just mentioned “Using the available architecture documentation and a series of modeling cycles ...”. Either the authors should not claim it as a contribution or add more details to portray their novelty in the approach.}

Using the available architecture documentation and a series of modelling cycles, we constrained the number of event counters used for the modelling to the set of the counters that have the most significant impact on the energy consumption and are suitable for static analysis. Most notably all these PMCs can be statically predicted from code-block size using architecture models, which makes them suitable for use in energy analysis tools. These counters also yield the highest observed estimation accuracy compared to physical measurements when compared with the retrieved estimations of other event counter combinations.
%; combinations of counters where there was no high-correlation between them and they were not always suitable for static analysis
The selected counters are shown in \Cref{tab:Cortex_M0_counters}. %Finally, we excluded the ``number of instruction reads from Flash memory'' counter (\emph{Flash\_insn\_reads}) from the modelling, because the ISS and the static energy analysis showed different fetch behaviour. Although this can have an impact on the accuracy of the energy consumption model, the extracted energy model is proven to be adequately accurate, as demonstrated in \Cref{subsec:Cortex_m0_model_training}.

\subsection{Model Training and Validation}
\label{subsec:Cortex_m0_model_training}

\begin{table*}[!htb]
\centering
\resizebox{\textwidth}{!}{%
\begin{tabular}{|l|l|c|c|}
\hline
\rowcolor[HTML]{C0C0C0} 
\multicolumn{1}{|c|}{\cellcolor[HTML]{C0C0C0}\textbf{Hardware Config.}} & \multicolumn{1}{c|}{\cellcolor[HTML]{C0C0C0}\textbf{Energy Consumption Model [nJ]}} & \textbf{Meas. Energy[J]} & \multicolumn{1}{c|}{\cellcolor[HTML]{C0C0C0}\textbf{MAPE [\%]}}  \\ \hline
[20, OFF, 0] & \begin{tabular}[c]{@{}l@{}} $\text{E} = 0.964258\times{C_{\text{1}}} + 1.652455\times{C_{\text{2}}} + 2.091986\times{C_{\text{3}}} + 1.109833\times{C_{\text{4}}} + 0.650563\times{C_{\text{5}}} + 0.633621\times{C_{\text{6}}}$ \end{tabular} & 221.4  & 2.80\\ \hline
[20, OFF, 1] & \begin{tabular}[c]{@{}l@{}} $\text{E} = 1.282474\times{C_{\text{1}}} + 2.110668\times{C_{\text{2}}} + 2.191545\times{C_{\text{3}}} + 1.185609\times{C_{\text{4}}} + 0.416602\times{C_{\text{5}}} + 1.178991\times{C_{\text{6}}}$ \end{tabular} & 274.9  & 2.97\\ \hline
[20, ON, 0] & \begin{tabular}[c]{@{}l@{}} $\text{E} = 1.003378\times{C_{\text{1}}} + 1.885309\times{C_{\text{2}}} + 1.802974\times{C_{\text{3}}} + 1.122833\times{C_{\text{4}}} + 0.849223\times{C_{\text{5}}} + 0.475831\times{C_{\text{6}}}$ \end{tabular} & 226.38 & 2.86\\ \hline
[20, ON, 1] & \begin{tabular}[c]{@{}l@{}} $\text{E} = 0.895879\times{C_{\text{1}}} + 2.185851\times{C_{\text{2}}} + 2.001178\times{C_{\text{3}}} + 1.493364\times{C_{\text{4}}} + 1.076354\times{C_{\text{5}}} + 1.573758\times{C_{\text{6}}}$ \end{tabular} & 227.9 & 3.68 \\ \hline
[24, OFF, 0] & \begin{tabular}[c]{@{}l@{}} $\text{E} = 0.959172\times{C_{\text{1}}} + 1.888565\times{C_{\text{2}}} + 1.357556\times{C_{\text{3}}} + 1.089427\times{C_{\text{4}}} + 0.993145\times{C_{\text{5}}} + 0.562952\times{C_{\text{6}}}$ \end{tabular} & 214.62 & 3.22 \\ \hline
[24, OFF, 1] & \begin{tabular}[c]{@{}l@{}} $\text{E} = 1.178558\times{C_{\text{1}}} + 2.540429\times{C_{\text{2}}} + 2.042475\times{C_{\text{3}}} + 1.190892\times{C_{\text{4}}} + 0.979651\times{C_{\text{5}}} + 0.891088\times{C_{\text{6}}}$ \end{tabular} & 264.88 & 3.16 \\ \hline
[24, ON, 0] & \begin{tabular}[c]{@{}l@{}} $\text{E} = 0.985415\times{C_{\text{1}}} + 1.933276\times{C_{\text{2}}} + 1.448160\times{C_{\text{3}}} + 1.075671\times{C_{\text{4}}} + 1.011891\times{C_{\text{5}}} + 0.617510\times{C_{\text{6}}}$ \end{tabular} & 220.03 & 3.36\\ \hline
[24, ON, 1] & \begin{tabular}[c]{@{}l@{}} $\text{E} = 0.883755\times{C_{\text{1}}} + 2.156046\times{C_{\text{2}}} + 1.633465\times{C_{\text{3}}} + 1.436556\times{C_{\text{4}}} + 1.152560\times{C_{\text{5}}} + 1.455166\times{C_{\text{6}}}$ \end{tabular} & 220.05 & 4.15\\ \hline
[48, OFF, 1] & \begin{tabular}[c]{@{}l@{}} $\text{E} = 1.096677\times{C_{\text{1}}} + 2.364495\times{C_{\text{2}}} + 1.627854\times{C_{\text{3}}} + 1.173680\times{C_{\text{4}}} + 0.681475\times{C_{\text{5}}} + 0.652665\times{C_{\text{6}}}$ \end{tabular} & 243.44 & 3.65 \\ \hline
[48, ON, 1] & \begin{tabular}[c]{@{}l@{}} $\text{E} = 0.816331\times{C_{\text{1}}} + 2.014612\times{C_{\text{2}}} + 1.372157\times{C_{\text{3}}} + 1.402116\times{C_{\text{4}}} + 0.835035\times{C_{\text{5}}} + 1.250446\times{C_{\text{6}}}$ \end{tabular} & 202.5 & 4.33\\ \hline
\end{tabular}%
}
\caption{Energy models for selected Cortex-M0 hardware configurations -- Hardware Configuration\ Format: [Frequency (MHz), PreFetch (ON/OFF), WaitState (0/1)] and MAPE: Mean Absolute Percentage Error}
\label{tab:models}
\end{table*}

%\begin{figure*}[!htp]
%\centering
%\resizebox{0.75\textwidth}{!}{%
%  \includegraphics[trim=0.5cm 12cm 6.2cm 1.6cm,clip=true,width=0.5\linewidth]{security_v2}
%}
%\caption{Relative costs associated with events used in the energy model at distinct processor configurations.}
%\label{fig:relativeEventWeigthAnalysis}
%\end{figure*}

When using regression modelling, it is critical to include as broad and representative a training sample as possible in the training phase. This ensures that the model is as generic as possible and can capture a large part of the space being modelled. Thus, instead of splitting our data into predefined training and testing sets, we included all data into the training, and we used k-fold cross-validation to ensure the retrieved models avoid over-fitting and selection bias. %If the cross-validation demonstrates a good estimation accuracy across all folds, then the final model using all the available data will exhibit a balanced variance and bias. Thus, the model will have a good chance of accurately capturing a big part of the space being modeled. 
In our case, we used 10-fold cross-validation and we used the $R^2$ to evaluate the performance of each of the ten models for each of the modelling configuration, shown in Table~\ref{tab:models}. 
%\todo[inline]{230 benchmarks (more than existing works) are used to improve the training and validation, which was performed with k-fold cross validation. However, analysis that evaluates the heterogeneity of these benchmarks would be useful, particularly as the model's ability to generalize is important. For example, 154 of the benchmarks are three different CNN layers with various hyper-parameters. Some of these benchmarks could potentially be very similar, which would undermine the k-cross fold validation, as workloads in the "unseen" set could have a corresponding highly similar workload in the training set. The R2 is naturally quite high when modelling this problem and perhaps the assumption regarding the generalization across the folds needs more thorough testing}
The 10-fold cross-validation yielded an $R^2$ mean value of close to 0.99 for all configurations, with a standard deviation of around 0.2\%, where an $R^2$ value close to 1 indicates an excellent prediction. This %is a robust result as the $R^2$ score approaches the value of one across all the different folds, 
demonstrates that the counters selected for the model are accurately capturing the energy consumption of a variety of programs. For the final model coefficients and results, all the data points were used in the training.

%\todo[inline]{How does the proposed model compare against state of the art (qualitative measures, in accuracy and cost)?}
Energy models for the different hardware configurations and their accuracy are listed in Table~\ref{tab:models}. For all models the MAPE is less than 5\%, %with a standard deviation of less than 5\%, 
compared to hardware energy measurements. Compared to other relevant works our models achieve lower error, while being trained and validated on a much larger variety of benchmarks using only statically predictable events suitable for code-block-level analysis~\cite{bazzaz2013accurate,ruberg2015microcontroller,konstantakos2008energy}.%. ~\cite{bazzaz2013accurate} achieve on average 3.65\% error across 7 benchmarks, ~\cite{ruberg2015microcontroller} achieve 2.81\% error on a single image processing workload and ~\cite{konstantakos2008energy} achieve 4.5\% average error on a single temperature sampling program.%An early version of the model, configured for predictabe  20 MHz frequency, PreFetch on, and WaitState 1 was evaluated in the context of static energy consumption estimation in~\cite{D4.4_TeamPlay} demonstrating the suitability of our models for static analysis.

Analysing the calculated model weights for the PMCs across the different hardware configurations shows a high variation, however some interesting general deductions can be made. For example, when the WaitState is 1 there is a higher cost associated with \textit{Flash reads}, due to the fact that the processor stays idle while waiting for data from the memory. Also, the  cost for \textit{RAM data reads} is close to or higher than \textit{RAM writes} and \textit{Flash reads}, because there are more than twice as many \textit{RAM data reads} operations than the other two and the NNLS solver associates a large part of the energy consumption to them, even if the operation itself uses much less energy. Introducing a WaitState clearly increases energy consumption and thus the PMC coefficients for the entire DUT (however it is needed for correct functionality at higher frequencies). When the WaitState is 0, turning on the PreFetch results in slightly higher energy consumption for the frequencies that support WaitState 0. Consequently, when the WaitState is 1 and PreFetch is ON, there is a significantly reduced overall energy consumption with lower model weights for arithmetic PMCs and branches, but higher model weights for data movement PMCs.%, however this ultimately results in a much lower energy consumption compared to when the PreFetch is off in all configurations.

\section{Conclusion and future work}

This paper offers an open-source, ready-to-use energy model for the Arm Cortex-M0 processor~\cite{discovery-data}. The model can be used for profiling-based analysis to accurately estimate the total energy consumption of a program and in static analysis to predict the energy budget of a particular block of code with a MAPE of less that 5\%.
The models also account for various frequency and flash instruction-buffer configurations of the processor that can significantly affect the execution time and energy consumption of an application. Our customised open-source ISS~\cite{Thumbulator} is also readily available to profile the execution time and energy consumption of edge computing applications for any of the STM32F0xx family of processors. This allows developers to choose the hardware configuration that can meet the resource requirements.
 
   \section*{Acknowledgement}

This research has been supported by the European Union's Horizon 2020 Research and Innovation Programme under grant agreement No. 779882, TeamPlay (Time, Energy and security Analysis for Multi/Many-core heterogeneous PLAtforms).

\begingroup
% \raggedright
% \sloppy
% \Urlmuskip=0mu plus 1mu\relax
%\todo[inline]{There are a lot of web references used. The authors might want to cut down on them and use the space for more relevant text.}
\bibliographystyle{alpha}
\bibliography{shortedRef}

\endgroup
\end{document}